# Dynamic Phase Transition in the Kinetic Spin-3/2 Blume-Capel Model: Phase Diagrams in the Temperature and Crystal-Field Interaction Plane

**Mustafa Keskin, Osman Canko, Bayram Deviren**
Department of Physics, Erciyes University, 38039 Kayseri, Turkey

---

**Abstract**

We analyze, within a mean-field approach, the stationary states of the kinetic spin-3/2 Blume-Capel model by the Glauber-type stochastic dynamics and subject to a time-dependent oscillating external magnetic field. The dynamic phase transition points are obtained by investigating the behavior of the dynamic magnetization as a function of temperature and as well as calculating the Liapunov exponent. Phase diagrams are constructed in the temperature and crystal-field interaction plane. We find five fundamental types of phase diagrams for the different values of the reduced magnetic field amplitude parameter (h) in which they present a disordered, two ordered phases and the coexistences phase regions. The phase diagrams also exhibit a dynamic double critical end point for $0<h\leq1.44$, one dynamic tricritical point for $1.44<h\leq5.06$ and two dynamic tricritical points for $h>5.06$.

---

The equilibrium aspects of cooperative physical systems are now rather well known [1] within the framework of equilibrium statistical physics. On the other hand, the nonequilibrium properties of the systems are not yet well known either theoretically or experimentally due to the complexity. Some interesting problems in nonequilibrium systems are the nonequilibrium or the dynamic phase transition (DPT) in which the mechanism behind it has not yet been explored rigorously and the basic phenomenology is still undeveloped. Hence, further efforts on these challenging time-dependent problems, especially calculating the DPT points and constructing the phase diagram, should promise to be rewarding in the future. The dynamic phase transition (DPT) in nonequilibrium system at the presence of an oscillating external magnetic field has attracted much attention recent years, theoretically [2]. Experimental evidences for the DPT have been found in magnetic systems [3]. In our recent paper [4], we have presented a study, within a mean-field approach, the stationary states of the kinetic spin-3/2 Blume-Capel model under a time-dependent oscillating external magnetic field. We use a Glauber-type stochastic dynamics to describe the time evolution of the system. We have investigated the behavior of the time-dependence of the magnetization and the behavior of the average magnetization in a period, also called the dynamic magnetization, as a function of reduced temperature. The nonequilibrium or dynamic phase transition (DPT) points are found by investigating the behavior of the dynamic magnetization as a function of the reduced temperature. These investigations are also checked and verified by calculating the Liapunov exponent. Finally, we presented the phase diagrams in the (T, h) plane. We found that the behavior of the system strongly depends on the values of d and six different phase diagram topologies, where the P, $F_{3/2}$, $F_{1/2}$, $F_{3/2}+F_{1/2}$, $F_{3/2}+P$, $F_{1/2}+P$ or/and $F_{3/2}+F_{1/2}+P$ phases occur that depend on the values of d, are obtained. The system also exhibits one or three tricritical points depending upon values of d. On the other hand, the phase diagrams of the kinetic spin-3/2 BC model under the presence of a time varying (sinusoidal) magnetic field in the temperature and crystal-field interaction plane were not presented by using the Glauber stochastic dynamics in Ref. 4. Therefore, the purpose of the present short communication is to present the phase diagrams of the kinetic spin-3/2 BC model under a time-dependent oscillating external magnetic field in the temperature and crystal-field interaction plane by using the Glauber stochastic dynamics [5].

Since the model and method are discussed in Ref. 4 extensively, we shall only give a brief summary in here. The spin-3/2 Ising system containing a single-ion anisotropy or crystal-field interaction (D) in addition to the bilinear exchanged interaction (J) is often called the spin-3/2 Blume-Capel (BC) model. The model has been one of the actively studied systems in the condensed matter and statistical physics because of a rich variety of critical and multicritical phenomena it exhibits.

The Hamiltonian of the spin-3/2 BC model is given by

$$H = -J\sum_{<ij>} S_i S_j - D\sum_i S_i^2 - H\sum_i S_i, \qquad (1)$$

where the $S_i$ located at the site $i$ on a discrete lattice can take the values $\pm 3/2$ and $\pm 1/2$ at each site $i$ of a lattice and $\langle ij \rangle$ indicates summation over all pairs of nearest-neighbor sites. The first term describes the ferromagnetic coupling (J>0) between the spins at sites $i$ and $j$. This interaction is restricted to z nearest-neighbor pairs of spins. The second term describes the crystal-field interaction or a single-ion anisotropy, and the last term, H, represents a time-dependent external oscillating magnetic field. H is given by $H(t)=H_0 \cos(wt)$, where $H_0$ and $w=2\pi v$ are the amplitude and the angular frequency of the oscillating field, respectively. The system is in contact with an isothermal heat bath at absolute temperature.

Since the system evolves according to a Glauber-type stochastic process at a rate of $1/\tau$ transitions per unit time. We define $P(S_1, S_2, \ldots, S_N; t)$ as the probability that the system has the S-spin configuration, $S_1, S_2, \ldots, S_N$, at time t. The time dependence of this probability function is assumed to be governed by the master equation which describes the interaction between spins and heat bath and can be written as

$$\frac{d}{dt} P(S_1, S_2, \ldots, S_N; t) = -\sum_i (\sum_{S_i \neq S_i'} W_i(S_i \to S_i')) P(S_1, S_2, \ldots, S_i, \ldots, S_N; t)$$
$$+ \sum_i (\sum_{S_i \neq S_i'} W_i(S_i' \to S_i) P(S_1, S_2, \ldots, S_i', \ldots, S_N; t)), \qquad (2)$$

where $W_i(S_i \to S_i')$, the probability per unit time that the *i*th spin changes from the value $S_i$ to $S_i'$. Since the system is in contact with a heat bath at absolute temperature T, each spin can change from the value $S_i$ to $S_i'$ with the probability per unit time

$$W_i(S_i \rightarrow S'_i) = \frac{1}{\tau} \frac{\exp(-\beta \Delta E(S_i \rightarrow S'_i))}{\sum_{S'_i} \exp(-\beta \Delta E(S_i \rightarrow S'_i))}, \qquad (3)$$

where $\beta = 1/k_B T$, $k_B$ is the Boltzmann factor, $\sum_{S'_i}$ is the sum over the four possible values of $S'_i$, $\pm 3/2$, $\pm 1/2$ and $\Delta E$ is the change in the energy of the system when the $S_i$-spin changes. The probabilities satisfy the detailed balance condition [4]. By using the Glauber-type stochastic dynamics with a mean-field approach, we obtain the mean-field dynamical equation for the magnetization [4].

$$\Omega \frac{d}{d\xi} m = -m + \frac{3\exp(d/T)\sinh[3(m+h\cos\xi)/2T] + \exp(-d/T)\sinh[(m+h\cos\xi)/2T]}{2\exp(d/T)\cosh[-3(m+h\cos\xi)/2T] + 2\exp(-d/T)\cosh[-(m+h\cos\xi)/2T]}, \qquad (4)$$

where $m \equiv \langle S \rangle$, $\xi = wt$, $T = (\beta z J)^{-1}$, $d = D/zJ$, $h = H_0/zJ$, and $\Omega = \tau w$. We fixed $z = 4$ and $\Omega = 2\pi$.

The stationary solution of Eq. (4) will be a periodic function of $\xi$ with period $2\pi$; that is, $m(\xi + 2\pi) = m(\xi)$. Moreover, it can be one of two types according to whether it has or does not have the property

$$m(\xi + \pi) = -m(\xi). \qquad (5)$$

A solution satisfying Eq. (5) is called a symmetric solution which corresponds to a paramagnetic (P) solution. In this solution, the magnetization $m(\xi)$ oscillates around the zero value and is delayed with respect to the external field. The second type of solution which does not satisfy Eq. (5) is called nonsymmetric solution that corresponds to a ferromagnetic (F) solution. In this case the magnetization does not follow the external magnetic field any more, but instead of oscillating around the zero value; it oscillates around a nonzero value, namely either $\pm 3/2$ or $\pm 1/2$. Hence, if it oscillates around $\pm 3/2$, this nonsymmetric solution corresponds to the ferromagnetic $\pm 3/2$ ($F_{3/2}$) phase and if it oscillates around $\pm 1/2$, this corresponds to the ferromagnetic $\pm 1/2$ ($F_{1/2}$) phase. These facts are seen explicitly by solving Eq. (4) numerically. The solutions and results are given and discussed in Fig. 1 of Ref. 4, extensively in which displays seven different phases, namely the P, $F_{1/2}$, $F_{3/2}$, $F_{1/2}+P$, $F_{3/2}+P$, $F_{3/2}+F_{1/2}$ and $F_{3/2}+F_{1/2}+P$ phases, exist in the system.

In order to obtain the dynamic phase boundaries among these phases, we have to calculate DPT points and then we can present phase diagrams of the system in the reduced temperature and crystal-field interaction plane. DPT points will be obtained by investigating the behavior of the average magnetization in a period, which is also called the dynamic magnetization, as a function of the reduced temperature

The average magnetization (M) in a period, namely the dynamic magnetization, is given as

$$M = \frac{1}{2\pi} \int_0^{2\pi} m(\xi) d\xi. \qquad (6)$$

The behavior of M as a function of the reduced temperature for several values of h and d are obtained by combining the numerical methods of Adams-Moulton predictor corrector with the Romberg integration and the behavior of M gives the DPT points and as well as type of the phase transition. We will obtain the DPT points and also the type of the phase transition from the behavior of M. For example, if M decreases to zero continuously as the reduced temperature increases, therefore a second-order phase transition occurs, $T_C$. If M decreases to zero discontinuously, a first-order phase transition occurs, $T_t$. Since we gave explanatory some interesting examples to illustrate the calculation of the DPT and the dynamic phase boundaries among ten phases in Fig. 2 of Ref. 4, we will not present the any behavior of M in this short communication. We should also mention that we calculated the Liapunov exponent to verify the stability of solutions and the DPT points in Ref. 4, seen in Fig. 3 of Ref. 4.

We can now obtain the phase diagrams of the system and the calculated phase diagrams in the (T, d) plane are presented in Figs. 1 (a)-(e). In these phase diagrams the solid and dashed lines represent the second- and first-order phase transition lines, respectively, the dynamic tricritical points are denoted by a solid circle, and the dynamic double critical end point is represented by B. As seen in Fig.1, we have obtained five different phase diagram topologies. (i) For 0<h≤0.33, we performed the phase diagram at h=0.125, seen in Fig. 1(a). The phase diagram displays the disordered or paramagnetic (P), the ferromagnetic-3/2 ($F_{3/2}$), the ferromagnetic-1/2 ($F_{1/2}$) and the $F_{3/2}+F_{1/2}$, also called coexistence phase or region. The dynamic phase boundaries among the P, the $F_{3/2}$ and $F_{1/2}$ phases are the second-order phase transition lines. For large negative values of d and low values of T, the $F_{1/2}$ phase exists and for high values of d and T, the $F_{3/2}$ phase occurs. Moreover, at low reduced temperatures, there is a range of values of d in which the $F_{3/2}$ and $F_{1/2}$ phases coexist, called the coexistence region or phase, $F_{3/2}+F_{1/2}$. The $F_{3/2}+F_{1/2}$ phase or region is separated from the $F_{3/2}$ and $F_{1/2}$ phases by the first-order phase lines. These two first-order phase lines start from zero temperature and terminate at the dynamic double critical end point B, where two critical phases coexist. The similar phase diagram without the $F_{3/2}+F_{1/2}$ phase has been also obtained by methods in the equilibrium statistical physics, namely the mean-field approximation and the Monte Carlo simulation [6], a renormalization-group transformation in position-space based on the Migdal-Kadanoff recursion relations [7] and in the exact solution of the model on Bethe lattice by using the exact recursion equation [8]. Moreover, the $F_{3/2}+F_{1/2}$ phase for low values of the reduced temperature (T) has been also found in the exact solution of the model on Bethe lattice by using the exact recursion equation [9]. (ii) For 0.33<h≤ 0.36, the phase diagram is constructed for h=0.35 and is similar to the phase diagram of Fig. 1(a), except for very low values of T three more coexistence phases, namely the $F_{1/2}$+P, $F_{3/2}$+P, $F_{3/2}+F_{1/2}$+P phases, occur, seen in Fig. 1(b). The dynamic phase boundaries among these coexistence phases are all first-order phase transition lines and the $F_{3/2}+F_{1/2}$ phase becomes very small. (iii) For 0.36<h≤1.265, we have presented the phase diagram for h=0.375, seen in Fig. 1(c). The phase diagram is similar to Fig. 1(b) but following differences have been found: (1) The second-order phase line and the $F_{1/2}$ phase occur at low temperatures disappear. (2) The dynamic double critical end point becomes the dynamic tricritical point where the both first-order phase transition lines merge and signals the change from the first- to the second-order phase transitions. (3) The $F_{3/2}+F_{1/2}$ phase or region does not exist any more. (iv) For 1.265<h≤1.35 the phase diagram is shown in Fig. 1(d) for h=1.3. This is the more interesting phase diagram in which the system exhibits two dynamic tricritical points besides the P, $F_{1/2}$, $F_{3/2}$+P and $F_{1/2}$ + P phases. The $F_{3/2}$ + P phase occurs for high values of d and two the $F_{1/2}$ + P phases exist for low values of d. The dynamic boundary between the P and $F_{1/2}$ phases is a second-order, but other boundaries among the other phases are all first-order lines. (v) For h>1.35, in this case the phase diagram was constructed for h=1.5, seen in Fig. 1 (e). The phase diagram is similar to the Fig. 1 (d), but only differs from Fig. 1 (d) in which the $F_{3/2}$+P phase does not exist any more.

Hence, the system exhibits two dynamic tricritical points and besides the P and the $F_{1/2}$ phases, only the $F_{1/2}+P$ coexistence region exist.

In conclusion, we found that the behavior of the system strongly depends on the values of h. Five fundamental types of phase diagrams, where the P, $F_{3/2}$, $F_{1/2}$, $F_{3/2}+P$, $F_{1/2}+P$, $F_{3/2}+F_{1/2}$ and/or $F_{3/2}+F_{1/2}+P$ phases occur that depend on values of h, are found. Moreover, the system also exhibits a double critical end point for $0<h\leq0.36$, one dynamic tricritical point for $0.36<h\leq1.265$ and two dynamic tricritical points for $h>1.265$. The stability of the solution and the DPT points are checked by calculating the Liapunov exponent. Finally, if one compares the phase diagrams in the (T, h) plane that presented in Fig. 4 of Ref. 4 and the phase diagrams in the (T, d) plane of this work, one can find following differences: (1) In the (T, h) plane, six fundamental phase diagrams are found, but in the (T, d) plane, five fundamental phase diagrams are obtained. (2) In the (T, h) plane, the system only exhibits one or three dynamic tricritical special points (one for $d\geq-0.4887$ and three for $d<-0.4887$), but in the (T, d) plane, a dynamic double critical end point for $0<h\leq0.36$, one dynamic tricritical point for $0.36<h\leq1.265$ and two dynamic tricritical points for $h>1.265$ are found. Hence, the system exhibits two dynamic special points, namely, the dynamic double critical end point and dynamic tricritical points. (3) Although the similar phase diagram of Fig. 1(a) without [6-8] and with [9] the $F_{3/2}+F_{1/2}$ phase has been also obtained by methods in the equilibrium statistical physics, we have not seen, best of our knowledge, any similar phase diagrams in the (T, h) plane of Ref. 4 in which have been calculated by methods in the equilibrium statistical physics.

**Acknowledgements**

This work was supported by the Scientific and Technological Research Council of Turkey (TÜBİTAK) Grant No. 105T114 and Erciyes University Research Funds, Grant No: FBA-06-01.

**List of the Figure Captions**

**Fig. 1.** Phase diagrams of the spin-3/2 Blume-Capel model in the (T, d) plane. Paramagnetic (P), Ferromagnetic-3/2 ($F_{3/2}$), Ferromagnetic-1/2 ($F_{1/2}$) phases, and four different coexistence phases or regions, namely the $F_{3/2}$+P, $F_{1/2}$+P, $F_{3/2}$+$F_{1/2}$ and $F_{3/2}$+$F_{1/2}$+P, are found. Dashed and solid lines represent the dynamic first- and second-order phase boundaries, respectively, the dynamic tricritical point is indicated with a solid circle and B is the dynamic double critical end point. **a)** h=0.125, **b)** h=0.35, **c)** h=0.375, **d)** h=1.3 and **e)** h=1.5.

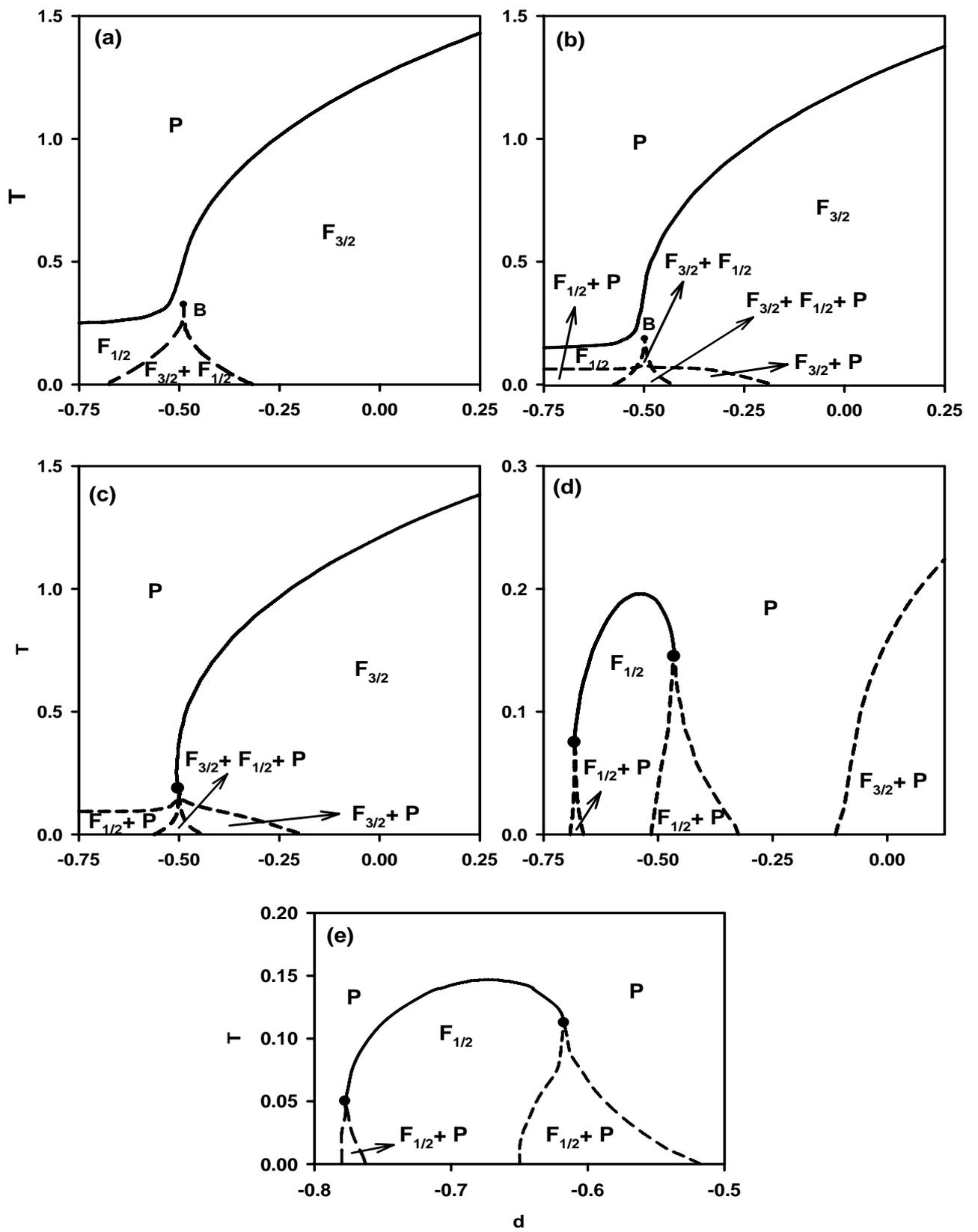

Fig. 1